# All-passive upconversion of incoherent near-infrared light at intensities down to ~$10^{-7}$ W/cm$^2$

Rabeeya Hamid[1], Demeng Feng[1], Pournima Narayanan[2], Justin S. Edwards[3], Manchen Hu[4], Emma Belliveau[4], Minjeong Kim[1], Sanket Deshpande[1], Chenghao Wan[4], Linda Pucurimay[5], David A. Czaplewski[6], Daniel N. Congreve[4], Mikhail A. Kats[1,3]*

[1]*Department of Electrical and Computer Engineering, University of Wisconsin-Madison, Madison, 53706, USA*

[2]*Department of Chemistry, Stanford University, Stanford, CA 94305, USA*

[3]*Department of Physics, University of Wisconsin-Madison, Madison, 53706, USA*

[4]*Department of Electrical Engineering, Stanford University, Stanford, CA 94305, USA*

[5]*Department of Materials Science and Engineering, Stanford University, Stanford, CA 94305, USA*

[6]*Center for Nanoscale Materials, Argonne National Laboratory, Argonne, Illinois 60439, USA*

*\*Corresponding author: mkats@wisc.edu*

## Abstract

Frequency upconversion, which converts low-energy photons into higher-energy ones, typically requires intense coherent illumination to drive nonlinear processes or the use of externally driven optoelectronic devices. Here, we demonstrate an upconversion system that converts low-intensity (down to ~$10^{-7}$ W/cm$^2$) incoherent near-infrared (NIR) light into the visible, reaching intensities perceptible by the human eye, without the use of any external power input. Our upconverting element is enabled by the following ingredients: (1) photon upconversion via triplet-triplet annihilation in a bulk heterojunction of the organic semiconductors Y6 and rubrene; (2) plasmonic enhancement of absorption and field intensity in the heterojunction layer; (3) collection enhancement using a dichroic thin-film assembly. To enable high-resolution imaging, the upconverting element is inserted at an intermediate image plane of a dual-wavelength telescope system, which preserves the relative directionality of rays between the incident NIR light and output visible light. Our all-passive upconversion imaging system will enable NIR imaging and sensing in low-light environments under energy constraints.

## 1. Introduction

Near-infrared (NIR) imaging is a crucial tool in a diverse set of scientific and industrial fields, including the biomedical industry[1–3], agricultural and environmental monitoring[4–6], food quality and safety assessment[7–10], facial recognition[11,12], and night vision[13–17]. Commercial NIR imaging systems are all externally powered, for example InGaAs cameras for industrial applications[18] and image intensifiers for night vision[19–21]. More recent work includes voltage-driven organic upconverting devices[22–24] and upconversion via wave mixing in nonlinear crystals[25–28]. The wave-mixing approach can function without external power, but only when operating at high input power densities (~MW/cm$^2$)[25,29,30].

Photon upconversion in semiconductors is an alternative that requires no external power input and can upconvert incoherent NIR light[31]. This process does not rely on conventional nonlinear wave mixing, but instead involves the sequential transfer of energy from two low-energy photons to two low-energy states to a single high-energy state, resulting in the emission of a high-energy photon. Among the best-established photon upconversion techniques are triplet-triplet annihilation (TTA)[32–36] and d- and f-subshell transitions in rare-earth-doped ions in solids[37–39]. TTA is particularly attractive for incoherent infrared imaging due to its relatively high efficiency at low-intensity irradiance and broadband absorption.

Here, we developed nanophotonic components that enhance the absorption and collection of light in the TTA process, and incorporated them into an optical system that preserves relative angles of input and output rays, enabling upconversion imaging with spatial resolution only limited by the aberrations of the optics. Our system enables all-passive upconversion detectable by the human eye, operating down to input NIR intensities around ~$10^{-7}$ W/cm$^2$. For context, such NIR intensities are roughly within an order of magnitude of the nightglow irradiance present in the night sky in the absence of the moon or other light sources[17] (Supporting Information, Section S13). The ability to passively upconvert low-intensity incoherent NIR light is an important step towards the development of all-passive and lightweight infrared imaging systems.

## 2. High-resolution upconversion imaging using triplet-triplet annihilation

We begin by exploring the imaging resolution and efficiency of an upconversion imaging system using state-of-the-art TTA materials without nanophotonic enhancement. We synthesized a ~100 - nm bulk heterojunction (BHJ) comprised of three organic semiconductors: Y6, rubrene, and tetraphenyldibenzoperiflanthene (DBP). Y6 functions as the sensitizer molecule, absorbing photons in the 700-930 nm range[32,35]. The excited states in Y6 then undergo several energy-transfer processes illustrated in Fig. 1a-c, resulting in the emission of a higher-energy photon from the DBP dopant dispersed in the annihilator—rubrene[32,35]. The resulting emission spectrum is centered on 610 nm.

Following TTA, the emitted visible light is incoherent (i.e., without a phase or angular relationship to the incident visible light) and hence we require an additional mechanism to preserve imaging through the upconversion process. To accomplish this, we position the upconverter BHJ at the mutual focal plane of a NIR imaging system on one side, and a visible imaging system on the other, reminiscent of a Keplerian telescope[40] (Fig. 1e). The imaging experiment is set up as shown in Fig. 1f. The NIR lens (L1) is achromatically corrected for the 700-930 nm absorption band of our Y6/rubrene/DBP BHJ. The resulting optical system enables upconversion imaging because the NIR lens converts the direction of the incident near-infrared rays to position at the upconverter, and then the position of the visible light emerging from the upconverter is converted back to direction of the output rays by the visible lens. Position on the BHJ is preserved during the upconversion process because the diffusion length of the excitons is small (< 40 nm[41,42]), which is far below the diffraction limit in our imaging system.

In Fig. 1g, we show that our upconversion imaging system resolves down to 100–110 line pairs per mm (lp/mm) on the upconverter under broadband ($\lambda = 750 - 1000$ nm) NIR illumination. This resolution level is comparable to that of commercial NIR viewing cameras that use image intensifiers, and surpasses other frequency upconversion techniques (at this field of view) in the literature[25,26,43,44]. the resolution in our experiment is primarily limited not by the upconverter film, but by chromatic aberration of the NIR lens (L1), which results in a focal length shift of around 10 µm across $\lambda = 750 – 930$ nm (Supporting Information, Section S10).

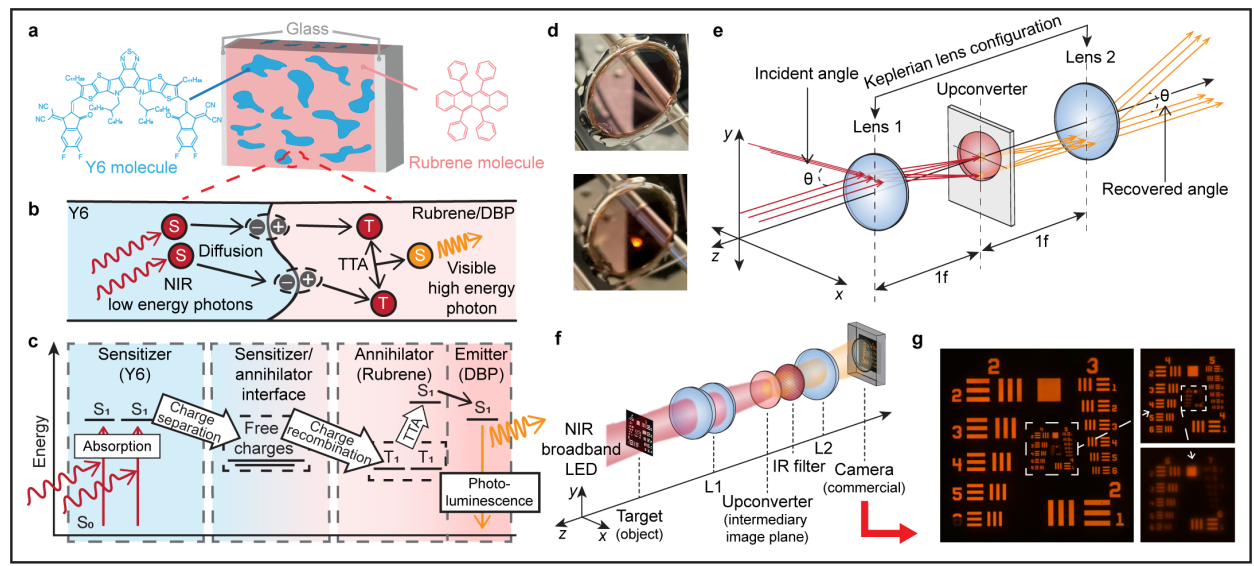

**Figure 1. NIR-to-visible upconversion imaging using a bulk heterojunction (BHJ) thin film. a**, The upconverting device consists of a BHJ of organic molecules Y6 and DBP-doped rubrene, sealed between two glass pieces using epoxy glue[32,33] to prevent oxygen ingress. **b-c**, Molecular diagram (b) and energy diagram (c) of TTA upconversion in the Y6/rubrene/DBP BHJ. Y6 absorbs incident NIR photons to generate excited singlets ($S_1$), which diffuse to the interface between Y6 and rubrene. At this interface, singlets transform into free charges and recombine to form charge-transfer states at the interface, which can then transfer to rubrene as molecular triplets ($T_1$). These triplets combine via TTA in rubrene to produce a high-energy singlet which transfers into DBP and then emits a high-energy photon via photoluminescence. **d**, Images of the upconverting thin film under ambient light are shown with (bottom) and without (top) NIR laser illumination. **e**, The upconverter is placed at the intermediate focal plane of the Keplerian lens system. This configuration preserves the relative angles between rays $\theta$ across the upconversion process, and therefore enables upconversion imaging. **f**, Schematic of the imaging setup used for transmission-mode imaging of an Air Force resolution target. **g**, Image captured using the setup in (f) with 1.17 magnification between the object and the image, using incoherent broadband illumination. Resolution groups 4 and higher are highlighted on the right.

## 3. Nanophotonic approaches to increase the power efficiency

Despite its ability to produce high-fidelity images under broadband incoherent illumination, the bare BHJ upconverter in Fig. 1 does not have the necessary efficiency (external quantum efficiency (EQE) of approximately 0.02-0.04%[35]) for many applications such as night-vision or telescopic eyepieces in the NIR. For these applications, we would like to convert NIR irradiance approaching night-time levels[17] to visible irradiance above the eye sensitivity threshold, which is in the range of 0.001 – 0.01 nW/cm$^2$, depending on the wavelength[45,46]. To reach the required system upconversion efficiency, we implemented dichroic thin-film stacks to enhance the collection of emitted visible light and plasmonic resonators to enhance the absorption of incident NIR light.

### 3.1: Increasing collection efficiency using a beaming dichroic backreflector

In our initial transmission imaging system (Fig. 1e-f), the roughly isotropic fluorescence of the Y6/rubrene/DBP BHJ results in the loss of more than 50% of the emitted visible light. To recover the backward-emitted light, we designed a thin-film coating (Fig. 2a-d), using alternate layers of the dielectrics niobium pentoxide ($Nb_2O_5$) and silicon dioxide ($SiO_2$), to act as a beaming dichroic backreflector (*Methods* 6.1). Our coating is transparent to incident NIR light and reflects visible light within a half angle of up to 30°—close to the critical emission angle from the BHJ to air. The total thickness of our coating was kept under 1 µm which, our simulations show, is small enough to prevent any degradation in imaging resolution due to lateral spread of the upconverted light via light propagation and reflection (Supporting Information, Section S3). Emission simulations (Supporting Information, Section S2) using the finite-difference time domain (FDTD) method reveal that we can expect an enhancement of approximately 2.3 at the peak emission wavelength (610 nm) of our upconverter. The enhancement beyond a factor of 2 is due to Purcell enhancement.

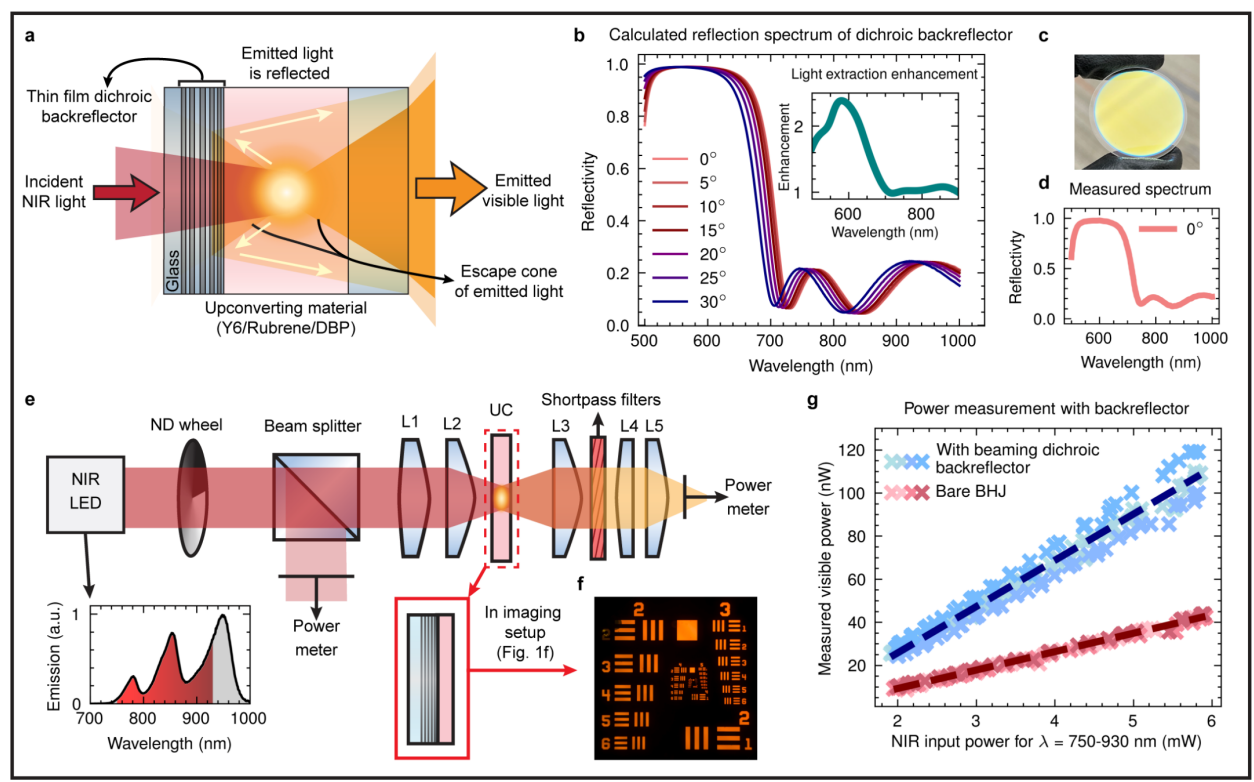

**Figure 2. Increasing collection efficiency using a beaming dichroic backreflector. a**, A thin-film dichroic backreflector allows NIR light to pass through while reflecting visible light. **b**, Calculated angle-dependent reflectance spectrum of the dichroic backreflector. Inset shows the simulated light extraction enhancement using FDTD simulations (Supporting Information, Section S2). **c**, Photo of the fabricated coating on a 1-inch diameter glass substrate. **d**, Measured spectrum of the fabricated backreflector shown in (c). **e**, Schematic of the power measurement setup using a broadband NIR source with three LEDs. Note that since the longer wavelengths in LED spectrum are not absorbed by the BHJ, we adjust the recorded input NIR power to remove the contribution of wavelengths longer than 930 nm. This setup is used to measure visible power from the upconverter in its original state (bare BHJ), and with the beaming dichroic backreflector. **f**, Image of the Air Force target captured using the setup in Fig. 1f with the dichroic backreflector. **g**, Visible power measured against NIR power for upconverters with and without the beaming dichroic backreflector. The different colors indicate either different upconverter samples or different spots on the same sample.

To experimentally validate the power enhancement, we constructed the transmission-mode power measurement setup in Fig. 2e. We employed a broadband NIR LED source and focused a 1 mm$^2$ circular beam onto the upconverter plane. We keep this incident beam area constant for all measurements because the TTA upconversion efficiency depends on the intensity of the incident light, due to the quadratic nature of TTA upconversion. It has been well-reported that the EQE increases with intensity and, beyond a certain threshold intensity, plateaus[32,35], reaching the linear power regime. Collimation optics and short-pass filters were used to capture the upconverted photoluminescence on the opposite end.

The output power from the bare BHJ (i.e., with glass on both sides), and with the dichroic backreflector, in the linear power regime is shown in Fig. 2g. The addition of the beaming dichroic backreflector leads to a 2.5-fold increase in the collected upconverted visible power. When we used these samples to perform an imaging experiment, we observed an increase in brightness without any apparent compromise in imaging quality or resolution (Fig. 2f vs Fig. 1g), as expected from simulations (Supporting Information, Section S3).

### 3.2: Absorption enhancement using localized plasmonic resonances

In a 100-nm film of Y6/rubrene/DBP BHJ, the NIR absorption is less than 25%, peaking around $\lambda = 800$ nm (Fig. 3b). While the film thickness can be increased further by increasing the Y6 concentration to increase the absorption, that actually decreases the internal quantum efficiency (IQE) and the overall EQE because it increases the probability of nonradiative quenching of singlet excitons in DBP-doped rubrene by nearby Y6 domains[35].

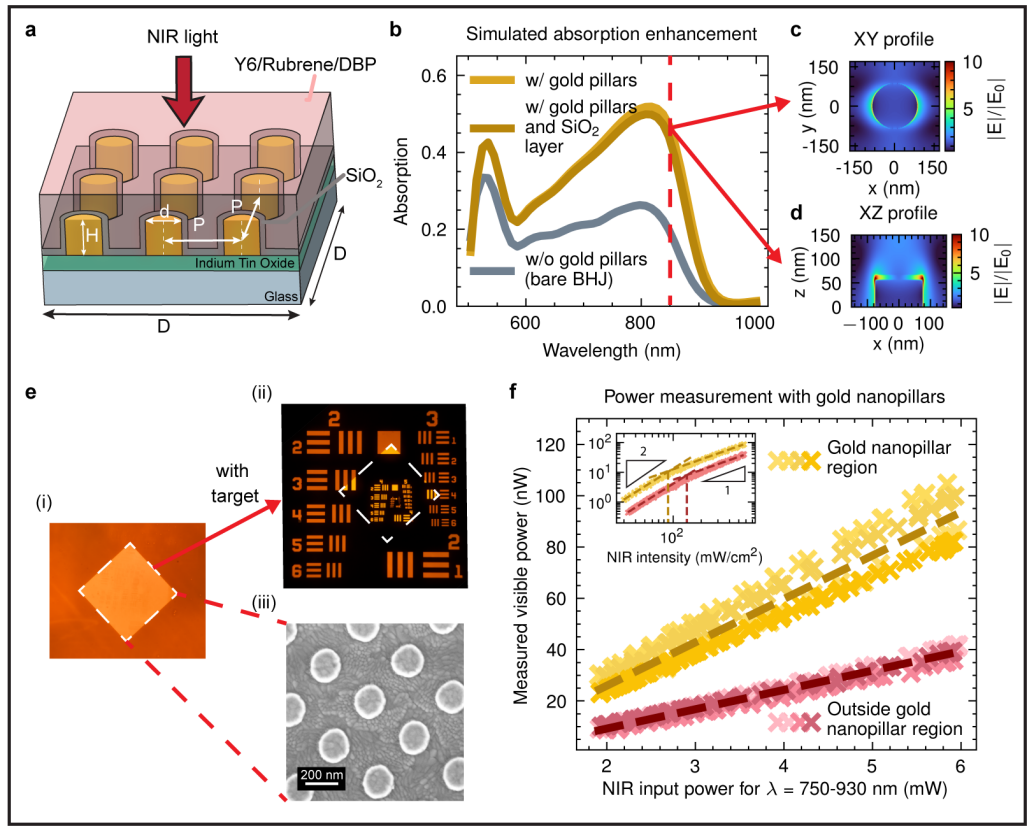

**Figure 3. Absorption enhancement using localized plasmonic resonances. a**, Periodic gold nanopillars are designed to be embedded in the upconverting material. A thin $SiO_2$ passivation layer is conformally deposited on the gold pillars to prevent direct contact with spin-coated Y6/rubrene/DBP and thus prevent electronic quenching. **b**, Simulated absorption using FDTD (Supporting Information, Section S4) of the upconverter in its original state, with periodic gold pillars, and with gold pillars and a 7 nm layer of $SiO_2$. **c-d**, Simulated field-enhancement profiles ($|E|/|E_0|$) across two cuts of the gold pillar at $\lambda = 850$ nm. **e**, (i), (ii) Images captured using the upconverter with gold nanopillar array (highlighted by a white square) in the Fig. 1f imaging setup with and without the Air Force target, (iii) Scanning electron microscope image of the fabricated gold nanopillar array. **f**, Measured visible power against input NIR power inside and outside the nanostructured region. The different colors indicate different samples or different spots on the same sample. The Inset shows the log-log plot, with the horizontal axis changed to NIR intensity on the upconverter. The intersection of the dotted lines is the estimated value of the quadratic-to-linear threshold.

Instead, we used localized plasmonic resonances to enhance NIR absorption by the BHJ in the 700-930 nm range. We designed periodically spaced gold nanopillars, as illustrated in Fig. 3a, with periodicity (P) of 340 nm, a diameter (d) of 170 nm, and a height (H) of 40 nm. Upon NIR excitation, the plasmonic resonances enhance the electric-field intensity in their proximity, increasing absorption in the BHJ. The nanopillar parameters were optimized to provide the highest absorption enhancement within the BHJ in the NIR range, while tolerating fabrication errors. Because the periodicity is smaller than half of the NIR wavelength, we do not expect any significant loss of imaging resolution. To avoid electronic quenching (Supporting Fig. S7f) at the BHJ/metal interface[47–50], a 7-nm dielectric ($SiO_2$) passivation layer was conformally deposited over the gold pillars before spin coating with Y6/rubrene/DBP. FDTD simulations (Fig. 3b and Supporting Information, Section S4) show an enhancement of ~2.1 across the NIR, with minimal impact from the passivation layer.

We used the same setup (Fig. 2e) to measure upconversion with and without the nanopillars on the same samples (Fig. 3f), observing a slope enhancement of 2.2 in the linear power regime, as expected from the simulation results. Imaging of the Air Force target shows that the integration of gold nanopillars does not adversely impact the imaging quality of the system (Fig. 3e (ii)). Though we did not optimize for it, the plasmonic resonators also reduce the quadratic-to-linear threshold of the TTA upconversion process, with a measured reduction of ~35% in the threshold NIR intensity across several nanostructured upconverting devices compared to the regions without nanopillars (Fig. 3f inset, and Supporting Information, Section S7). This effect is consistent with a previous report showing the reduction of the TTA threshold for organic upconverting devices using propagating surface plasmons[51].

## 4. High-efficiency upconversion imaging with fully integrated nanophotonic upconverter

The fully integrated upconverter combines the visible beaming (Fig. 2g) and NIR absorption enhancement (Fig. 3f) into a single device shown in Fig. 4, which has the beaming dichroic backreflector on one side and the nanopillar array on the other side, with the Y6/rubrene/DBP BHJ in the middle. The measured upconversion power is shown in Fig. 4a, with the backreflector contributing an enhancement of ~2.2 and the plasmonic nanopillars providing another factor of ~1.8, leading to an overall enhancement of ~4. This is slightly lower than what we initially expected from FDTD simulations (~4.6), and can be explained by the presence of a ~0.25 μm air gap in the sandwiched structure (Fig. 4a, details in Supporting Information, Section S9), which is hard to avoid with the current assembly method.

Using this fully integrated upconverter, we are able to upconvert NIR light at very low intensities, as shown in Fig. 4b. We observe that the fully integrated upconverter is emitting around 2 nW/cm$^2$ of visible light (output beam area: 0.005 cm$^2$) even at our lowest NIR illumination intensity, which is around 50 nW/cm$^2$ (using an input beam area of ~4 cm$^2$). This input intensity is orders of magnitude lower than what has been demonstrated in upconversion imaging systems based on nonlinear wave mixing[30] or lanthanide-doped nanoparticles[52], and is an order of magnitude lower than a recent demonstration using organic upconverting devices that used an external voltage source[24] (see Table S2 for a detailed comparison). An imaging system using the same NIR optics and upconverter designed for human-eye viewing would output a collimated visible beam (assuming a collimated input) with area equal to the dilated pupil of a human eye, which is ~0.5 cm$^2$; under these assumptions, the output intensity would be ~0.02 nW/cm$^2$, which is in the range of scotopic (dark-adapted) vision (see Supporting Information, Section S6 for a detailed calculation).

Note that, in Fig. 4b, the measured output power for the lowest NIR intensities is actually slightly higher from just bare glass (no upconverter, control) than the case of the bare BHJ with no nanostructures. This is an artifact of our measurement setup, where we use short-pass filters to block any transmitted NIR light from reaching our power meter, but some leaks through. In addition, there is some ambient background radiation captured by the power meter and the dark current noise, which also contributes to the baseline reading on the power meter (about ~38 nW/cm$^2$ in Fig 4b).

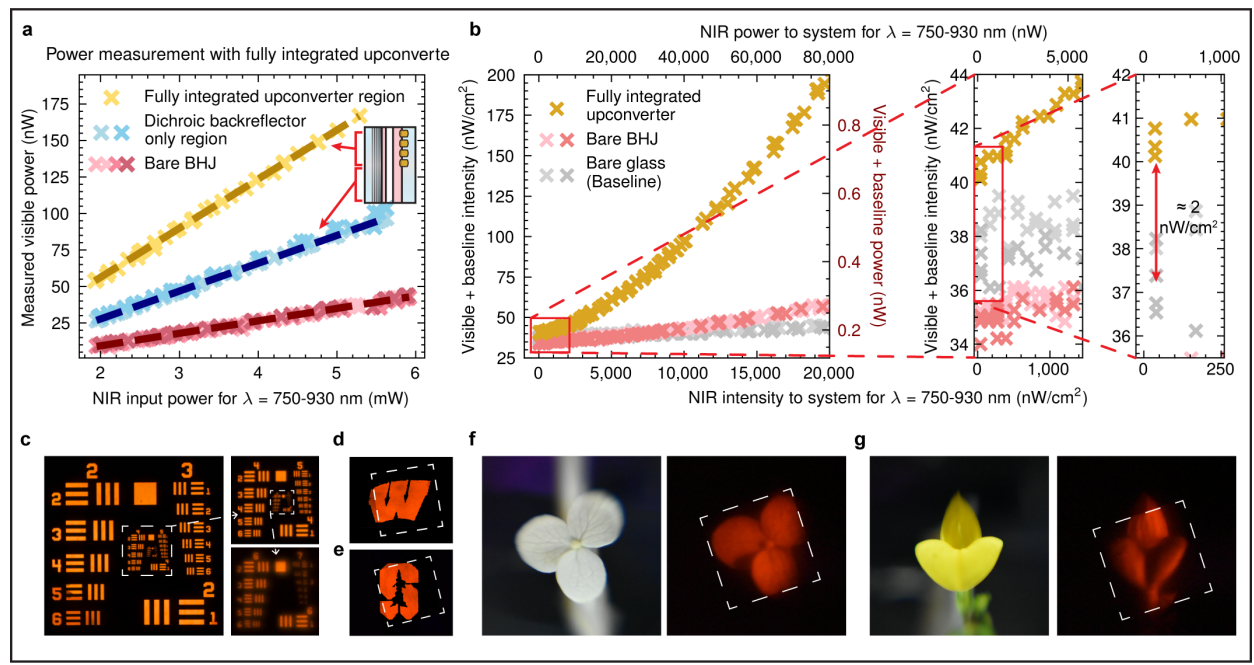

**Figure 4. Power measurement and imaging using the fully integrated upconverter. a**, Output visible power measured against input NIR power for different regions on the fully integrated upconverter, compared to a bare BHJ upconverter. **b**, NIR intensity to system (beam area of ~4 cm$^2$), as measured before focusing

onto the upconverter (beam area of ~0.004 cm$^2$), against the visible (+ baseline) intensity measured on power meter. Closeups on the right show the region of lowest input intensity. Details of lens system and beam sizes are given in Supporting Information, Section S5. **c**, Photos captured using the imaging system in Fig. 1f using the fully integrated upconverter with a (6 mm)$^2$ gold nanopillar array. Zoomed-in insets on the right show that the resolution of the system is approaching 100 line pairs per mm (more details in Supporting Information, Section S10). **d-e**, Images captured with the same setup as (c), but with the Air Force target replaced with paper cut-outs of the University of Wisconsin-Madison and Stanford University logos. **f-g**, Images captured with diffuse NIR light (~10 mW/cm$^2$) illuminating **f)** white hortensia, and **g)** bird's-foot trefoil flowers, using the fully integrated upconverter with the (6 mm)$^2$ array of plasmonic resonators (as highlighted by a white square).

For the final imaging experiments, we fabricated a 6-mm square array of plasmonic resonators to use in the fully integrated upconverter. We experimentally verified that there is no loss in resolution due to the simultaneous integration of the plasmonic structures and beaming dichroic backreflector by imaging the Air Force target (see Supporting Information, Section S10). In Fig. 4f-g, we also show upconversion imaging of small flowers, illuminated by roughly ~10 mW of diffused NIR light in absorption range of Y6, highlighting the ability of our upconversion system to image scattering objects illuminated by incoherent low-intensity NIR light.

## 5. Conclusion

In this paper, we experimentally realized an integrated upconversion system that can take incoherent near-infrared (NIR) light at intensities as low as ~$10^{-7}$ W/cm$^2$ and upconvert it into the visible, attaining intensities within human-eye sensitivity. Our all-passive imaging system requires no input power, either electrical or via optical pumping, and works with incoherent NIR input light over the range of 700 – 930 nm. These performance metrics are attained via a combination of a high-efficiency triplet-triplet annihilation (TTA) upconversion process within an organic bulk heterojunction, nanophotonic integrations that enhance the absorption using resonant localized surface plasmons, and the enhancement of visible light collection efficiency using a beaming dichroic backreflector. This engineered upconversion element is positioned within an imaging system that leads to a preservation of relative incident angles between incident and upconverted rays, therefore enabling high-resolution imaging, limited by the aberrations of our optics rather than by the upconverting element.

We envision all-passive high-efficiency upconversion optics as enabling applications from night vision to astronomical observation to NIR inspection in industrial and agricultural settings.

## Acknowledgements

We acknowledge funding from the Defense Advanced Research Projects Agency (DARPA) grant HR00112220010. Work performed at the Center for Nanoscale Materials, a U.S. Department of Energy Office (DoE) of Science User Facility, was supported by the U.S. DOE, Office of Basic Energy Sciences, under Contract No. DE-AC02-06CH11357. The authors also gratefully acknowledge the use of facilities and instrumentation at the UW-Madison Wisconsin Centers for Nanoscale Technology (wcnt.wisc.edu) partially supported by the NSF through the University of Wisconsin Materials Research Science and Engineering Center (DMR-2309000). PN acknowledges the support of a Stanford Graduate Fellowship in Science & Engineering (SGF) as a Gabilan Fellow and the Chevron Energy Fellowship. JSE was partially supported by the Department of Defense (DoD) through the National Defense Science & Engineering Graduate (NDSEG) Fellowship Program. LP acknowledges support of a National Science Foundation (NSF) Graduate Research Fellowship under grant DGE-2146755.

## Conflict of interest

DNC is a co-founder of and Chief Scientific Advisor to Quadratic 3D, Inc. Several of the authors have a patent pending on some of the techniques described in this paper.

## 6. Methods

### 6.1: Design and fabrication

**Design and deposition of the dichroic backreflector:**

To design a thin film coating using $Nb_2O_5$ and $SiO_2$, we first implemented the transfer matrix method in MATLAB. We choose an arbitrary number of alternate layers N and initialized the thicknesses $d_i$ randomly. We propagate a plane wave through the stack [air, glass, thin-film, BHJ, glass, air] and maximize the reflection at the BHJ/glass interface in the visible region (500-650 nm), and also maximize the transmission of NIR light (750-900 nm) into the BHJ. We used a regularization parameter ρ to balance between the two objectives and create a final objective function which is then fed into the MATLAB nonlinear optimization function `'optimproblem'`.

The optimization problem was run for several values of N, $d_i$, and ρ, with angle of incidence ($\theta_{opt}$) set to 20°. To maintain high imaging resolution, we aimed to keep our total thickness below 1 μm and hence we settled on using a total of N = 12 total layers for our design. The resulting design was tweaked using the regularization parameter until we saw satisfactory performance across both NIR and visible wavelengths. The final thicknesses are given by: [61, 69, 69, 103, 62, 107, 60, 104, 60, 100, 62, 55] nm, where the first layer is $Nb_2O_5$, the second is $SiO_2$, and so forth. The Y6/rubrene/DBP BHJ is positioned after the last layer in this stack. This design was provided to the thin film foundry Lohnstar Optics and was deposited on 1-mm thick 1" diameter B-270 glass substrates using DC magnetron sputtering.

**Design and fabrication of gold nanopillars:**

The gold nanopillars were designed to have a broadband plasmonic resonance in the NIR range centered around 850 nm. To find the optimal parameters, we swept across radii (50-110 nm), period (300-450 nm), and height (20-60 nm) of pillars and measured absorption inside Y6/rubrene/DBP BHJ using FDTD simulations (Supporting Information, Section S4). We use the material properties inside the simulation in order to distinguish metal from the BHJ and calculate the absorption inside the BHJ only. These absorption enhancement profiles were analyzed for all of the sweeps, and we chose the parameters that provided the highest broadband absorption enhancement and stable performance vs slight changes in the parameters, to account for fabrication errors.

We fabricated these gold pillars on indium-tin-oxide (ITO) coated soda lime glass squares (MSE Supplies, 1.1 mm thick, 12-15 Ohm/Sq, 15- or 25-mm length) using electron-beam lithography. The ITO layer prevented charging during the electron-beam write. We first cleaned the substrates by sonicating in acetone and then isopropanol (IPA). Then we spun two layers of resist on the substrates: PMMA 495K A3 at 2000 rpm, 60 sec, 500 rpm/sec and baked at 180°C for 3 minutes, and then PMMA 950K A6 at 4000 rpm, 60 sec, 500 rpm/sec, also baked at 180°C for 3 minutes. These samples were exposed using a JEOL 8100FS with current 4 nA and a total grid length of D, where D was either 2, 4, or 6 mm. The exposed samples were developed by sonicating in a 1:3 ratio of MIBK:IPA followed by plasma descum in the RIE March CS-1701 chamber. Gold was deposited using electron beam evaporation (Temescal FC2000), with a 5 nm adhesion layer of titanium. These samples were then soaked in 99% anisole at 70 °C overnight, and liftoff was completed by sonicating in clean anisole and then acetone. The last step was the $SiO_2$ passivation layer, which was deposited using plasma-enhanced chemical vapor deposition (PECVD). We approximate the thickness of $SiO_2$ to be close to 7 nm, as measured on homogenous gold thin films using variable-angle spectroscopic ellipsometry.

### 6.2: Synthesis

**Preparation of the Y6/Rubrene/DBP BHJ:**

Materials: Rubrene (>99.0%, purified by sublimation) was purchased from TCI America, DBP (5,10,15,20-Tetraphenylbisbenz[5,6]indeno[1,2,3-cd:1′,2′,3′-lm]perylene, >99.0%, purified by sublimation) and Y6 ( >99.0%) were purchased from Luminescence Technology Corp.

The upconversion components were dissolved in chloroform (2.5 mg/mL Y6, 14 mg/mL rubrene, and 0.2 mg/mL DBP) and filtered using a 0.2 -micron PTFE filter. All glass substrates were washed by sonicating for 10 minutes in a 1% Hellmanex detergent solution in deionized water, followed by two 5-minutes sonications in deionized water, followed by two 5-minute sonications in acetone, and two 5-minute sonications in isopropanol. The substrates were then dried under pressurized air to remove solvent and dust particles.

All substrates and encapsulation pieces were first treated for 15 minutes in UV-ozone. Then these were brought inside a nitrogen-filed glovebox, where Y6/rubrene/DBP solution was spun on top of the substrate at 1000 rpm for 30 seconds, with a 1666 rpm/second ramp. The coated substrates were then annealed at 70 °C for 5 minutes. The encapsulation piece was placed on top, and all sides of the upconverting material were thoroughly encapsulated using UV-curing epoxy (Norland Optical NOA81 from Thorlabs). Rubrene is especially sensitive to contact with oxygen and requires utmost care to prevent degradation.

For the preparation of bare BHJ samples, we spin on 1-mm thick glass substrates (soda-lime, B-270) and encapsulate using a standard 1-mm thick microscope slide. For the BHJ on top of the dichroic backreflector, we spin-coated the BHJ directly on top of the thin-film stack and encapsulated with a standard microscope slide. For the samples with plasmonic resonators, we spin-coated the BHJ on top of a substrate patterned with gold nanopillars and encapsulated with a standard microscope slide. For the final fully integrated upconverter, we spun BHJ film of half-thickness (using half of the concentration) on the thin-film coated substrate and full-thickness on the gold nanopillar substrate, and then pressed the two sides together (see also Supporting Information, Section S9). All substrates are 1-inch in length/diameter or smaller.

### 6.3: Imaging and power measurements

For all imaging and power measurements we use a NIR broadband LED from Thorlabs (MBB2L1) with three peaks at 770 nm, 860 nm, and 940 nm. We also use a combination of a 700 nm long-pass filter (Thorlabs, FELH0700) and a 1000 nm short-pass filter (FESH1000) to crop the wavelength region to 700 – 1000 nm. We collimated and spatially filtered the LED light to get a uniform circular beam for the power measurements. In addition – since the longer wavelengths in the LED spectrum are not absorbed by the BHJ, we adjust our recorded input power to get rid of the contribution of power for wavelengths longer than 930 nm. Since the LED starts emitting at 750 nm, the effective input NIR power for all measurements is in the range of 750 – 930 nm.

**Transmission imaging setup:**

The setup is shown in Fig. 1f. The NIR light passes through a negative USAF resolution test target (Thorlabs, R1DS1N). We estimate that the input beam to the imaging system is in the order of ~30 mW for Fig. 1g, Fig. 2f, and Fig. 3e (i)-(ii) and we reduce it down to ~20 mW for images with fully integrated upconverter in Fig. 4c. An achromatic 1:1 lens pair (L1 in Fig. 1f.) with an effective focal length of 100 mm (Edmund Optics, 47-302) is used to form an intermediary image plane on the upconverter. L1 is placed such that it is a single focal length away from both the target and upconverter and adjusted slightly to bring the target into focus on the camera. The image on the upconverter is magnified by ~1.17. We use a short-pass filter with cutoff wavelength 700 nm (Thorlabs, FESH0700) to filter out the NIR light from our upconverted light. For the imaging lens (L2 in Fig. 1f.), we use a plano-convex lens with a focal length of 100 mm (Edmund Optics, 38-321), and this lens is adjusted to provide the best focus and largest magnification on the digital camera (Nikon D5600). This camera was used together with the Laowa 100mm f/2.8 2X Ultra Macro APO lens which was set to the closest focus with f/2.8 and ISO 100. The shutter speed is varied across different images. For Fig. 1g a shutter speed of 30 seconds was used, for Fig. 2f and Fig. 3e. (i)-(ii), it was set to 20 seconds, and the Fig. 4c image was taken at 25 seconds. For the imaging resolution, refer to Supporting Information, Section S10.

**Scattered-light imaging setup:**

For the flower images shown in Fig. 4f-g, the imaging setup was similar to the transmission setup. NIR light from the LED (at around 30 mW) is passed through a 1500 grit diffuser (Thorlabs, DG10-1500-B) and used to illuminate the flower at around an angle of 45 degrees. We estimate that the NIR power illuminating the flower is in the order of ~10 mW. We then use a 20 mm achromatic aspheric lens (Edmund Optics, 85-301) to focus the scattered light from the flower onto the fully integrated upconverter. The distance between the flower and lens, and the lens and the upconverter are all nearly 20 mm (focal length of the lens) for 1:1 imaging and adjusted very slightly to bring the flower into focus. After the upconverter, we used a short-pass filter with cutoff wavelength 700 nm (Thorlabs, FESH0700) to filter out the NIR light from our upconverted light. Then we use plano-convex lens with a focal length of 50.8 mm (Edmund Optics, 62-601) for imaging on our digital camera. We use a 50 mm f/1.4D AF NIKKOR lens, which is set to f/5.6 and ISO 8000 for the images in Fig. 4f-g. For Fig. 4f, left to right, the shutter speeds used are: 1/200 seconds and 4 seconds, and for Fig. 4g they are: 1/500 seconds and 6 seconds.

**Power measurement setup:**

For the power measurement setup in Fig. 2e, we use a continuous ND filter wheel (Thorlabs, NDM2) and a 10:90 (R:T) non-polarizing beam splitter (Thorlabs, BS041) to control and measure the input power to the upconverter. The 10% reflected beam is measured using a silicon power meter (Thorlabs, S130VC) – we will call this $P_{BS}$. L1 (Thorlabs, ACA254-030-B) and L2 (Thorlabs, ACL25416U-B) are lenses used to focus light on the upconverter to 1 mm$^2$ spot size. L3 is a high-NA aspheric lens (Thorlabs, ACL25416U-A) used to capture a large cone of visible light from the upconverter and collimate it. We use a stack of NIR filters to completely eliminate NIR light before the power meter: three short-pass filters (Thorlabs, FESH700) and one NIR absorbing filter (Newport, FSR-KG3). The remaining visible light is focused on the power meter (Thorlabs, S130C, set to 610 nm detection, $P_{VIS}$) using 25.4 mm focal length planoconvex lenses L4 (Edmund Optics, 62-599) and L5 (Thorlabs, LA1951-AB). This optical set-up is also shown in Fig. S5c.

For all power measurements, we rotate the ND wheel and record $P_{BS}$ and $P_{VIS}$ for all upconverting samples. Then we replace the upconverter with a thermal power sensor (Thorlabs, S401C) and again rotate the ND wheel and record $P_{BS}$ and the input power to the upconverter– which will be referred to as $P_{NIR}$. $P_{NIR}$ and $P_{BS}$ follow a linear relationship, which is then used to map the $P_{BS}$ to $P_{NIR}$ in the datasets recorded for upconverting samples. We then multiply $P_{NIR}$ by a correction factor in order to eliminate the contribution of wavelengths longer than 930 nm.

In addition, we also replace the upconverter with glass and record the visible power ($P_{BG}$) and subtract it from $P_{VIS}$ in order to eliminate background noise or any NIR leaking through filters. Note that all power measurements took place with very low ambient light and under blackout fabric (Thorlabs, BK5).

**Minimum intensity measurement setup:**

For the intensity-intensity plot shown in Fig. 4b we use the same power measurement system as Fig. 2e. However, to measure upconversion yield at very low intensities (~$10^{-7}$ W/cm$^2$), we allowed the beam to diverge after the beamsplitter until it was nearly 4 cm$^2$ in area, then focused it down to nearly 0.004 cm$^2$ on the upconverter (shown in Supporting Information, Section S5). We decreased the input power by adding ND filters before the beam splitter, right until we were able to get the lowest measurable signal on the thermal power sensor (Thorlabs, S401C) which corresponded to minimum light intensity approaching 40 nW/cm$^2$. Following the same method as before, we did several power measurements with bare glass, bare BHJ, and our fully integrated device. Note that we do not subtract $P_{BG}$ from $P_{VIS}$, instead it is plotted alongside $P_{VIS}$ data (bare glass, baseline reading) in Fig. 4b. We also measured the beam spot at the power meter plane to be nearly 0.005 cm$^2$. All beam spot measurements were done using a monochrome CMOS camera (Thorlabs CS165MU1).

28. Valencia Molina, L., Camacho Morales, R., Zhang, J., Schiek, R., Staude, I., Sukhorukov, A. A. & Neshev, D. N. Enhanced Infrared Vision by Nonlinear Up-Conversion in Nonlocal Metasurfaces. *Advanced Materials* **n/a**, 2402777.

29. Liu, H., Li, H., Zheng, Y. & Chen, X. Nonlinear frequency conversion and manipulation of vector beams. *Opt. Lett., OL* **43**, 5981–5984 (2018).

30. Chalopin, B., Chiummo, A., Fabre, C., Maître, A. & Treps, N. Frequency doubling of low power images using a self-imaging cavity. *Opt. Express, OE* **18**, 8033–8042 (2010).

31. Ye, C., Zhou, L., Wang, X. & Liang, Z. Photon upconversion: from two-photon absorption (TPA) to triplet–triplet annihilation (TTA). *Physical Chemistry Chemical Physics* **18**, 10818–10835 (2016).

32. Izawa, S. & Hiramoto, M. Efficient solid-state photon upconversion enabled by triplet formation at an organic semiconductor interface. *Nat. Photon.* **15**, 895–900 (2021).

33. Wu, M., Congreve, D. N., Wilson, M. W. B., Jean, J., Geva, N., Welborn, M., Van Voorhis, T., Bulović, V., Bawendi, M. G. & Baldo, M. A. Solid-state infrared-to-visible upconversion sensitized by colloidal nanocrystals. *Nature Photon* **10**, 31–34 (2016).

34. Singh-Rachford, T. N. & Castellano, F. N. Photon upconversion based on sensitized triplet–triplet annihilation. *Coordination Chemistry Reviews* **254**, 2560–2573 (2010).

35. Hu, M., Belliveau, E., Wu, Y., Narayanan, P., Feng, D., Hamid, R., Murrietta, N., Ahmed, G. H., Kats, M. A. & Congreve, D. N. Bulk Heterojunction Upconversion Thin Films Fabricated via One-Step Solution Deposition. *ACS Nano* **17**, 22642–22655 (2023).

36. Narayanan, P., Hu, M., Gallegos, A. O., Pucurimay, L., Zhou, Q., Belliveau, E., Ahmed, G. H., Fernández, S., Michaels, W., Murrietta, N., Mutatu, V. E., Feng, D., Hamid, R., Yap, Muk Kam, K., Schloemer, T. H., Jaramillo, T. F., Kats, M. A. & Congreve, D. N. Overcoming the Absorption Bottleneck for Solid-State Infrared-to-Visible Upconversion. Preprint at https://doi.org/10.26434/chemrxiv-2024-h0k05 (2024).

37. Ye, S., Song, E.-H. & Zhang, Q.-Y. Transition Metal-Involved Photon Upconversion. *Advanced Science* **3**, 1600302 (2016).

38. Auzel, F. Upconversion and Anti-Stokes Processes with f and d Ions in Solids. *Chem. Rev.* **104**, 139–174 (2004).

39. Tiwari, S. P., Kumar, K. & Rai, V. K. Latent fingermarks detection for La2O3:Er3+/Yb3+ phosphor material in upconversion emission mode: A comparative study. *Journal of Applied Physics* **118**, 183109 (2015).

40. Salman, J., Gangishetty, M. K., Rubio-Perez, B. E., Feng, D., Yu, Z., Yang, Z., Wan, C., Frising, M., Shahsafi, A., Congreve, D. N. & Kats, M. A. Passive frequency conversion of ultraviolet images into the visible using perovskite nanocrystals. *J. Opt.* **23**, 054001 (2021).
15

Supporting information:

# All-passive upconversion of incoherent near-infrared light at intensities down to ~$10^{-7}$ W/cm$^2$

Rabeeya Hamid[1], Demeng Feng[1], Pournima Narayanan[2], Justin S. Edwards[3], Manchen Hu[4], Emma Belliveau[4], Minjeong Kim[1], Sanket Deshpande[1], Chenghao Wan[4], Linda Pucurimay[5], David A. Czaplewski[6], Daniel N. Congreve[4], Mikhail A. Kats[1,3]*

[1]Department of Electrical and Computer Engineering, University of Wisconsin-Madison, Madison, 53706, USA
[2]Department of Chemistry, Stanford University, Stanford, CA 94305, USA
[3]Department of Physics, University of Wisconsin-Madison, Madison, 53706, USA
[4]Department of Electrical Engineering, Stanford University, Stanford, CA 94305, USA
[5]Department of Materials Science and Engineering, Stanford University, Stanford, CA 94305, USA
[6]Center for Nanoscale Materials, Argonne National Laboratory, Argonne, Illinois 60439, USA

*Corresponding author: mkats@wisc.edu

## S1. Characterization of thickness and refractive index of Y6/rubrene/DBP bulk heterojunction (BHJ) via spectroscopic ellipsometry

We performed spectroscopic ellipsometry measurements on Y6/rubrene/DBP samples over a wavelength range of 500–1200 nm, using a J. A. Woollam V-VASE ellipsometer.

Because rubrene is prone to quick oxidation in air[1], leading to the change of the refractive index, we placed the sample into a dry $N_2$ chamber (Linkam THMSEL350V) during measurements. We sandblasted the backside of the substrate to minimize back reflection, so that the sample could be modeled as an isolated Y6/rubrene/DBP BHJ film on top of an infinitely thick substrate, with air on top.

We used a seven-Gaussian-oscillator model to fit the ellipsometry data, using the following form of Gaussian oscillators, as a function of photon energy[2]:

$$\epsilon_r(E) = \epsilon_1(E) + i\epsilon_2(E) \tag{S1}$$

$$\epsilon_2(E) = Ae^{-\left(\frac{E-E_0}{\sigma}\right)^2} - Ae^{-\left(\frac{E+E_0}{\sigma}\right)^2}, \tag{S2}$$

$$\epsilon_1(E) = \frac{2}{\pi} \mathcal{P} \int_0^\infty \xi \frac{\epsilon_2(\xi)}{\xi^2 - E^2} d\xi, \tag{S3}$$

$$\sigma = \frac{Br}{2\sqrt{\ln(2)}}, \tag{S4}$$

where $\epsilon_1$ is the real part of the relative permittivity $\epsilon_r$, $\epsilon_2$ is the imaginary part of the relative permittivity, $A$ is the amplitude of the oscillator (dimensionless), $Br$ is the broadening of the oscillator (eV), $E_0$ is the center energy of the oscillator (eV), and $E$ is the photon energy (eV). $\mathcal{P}$ is the Cauchy principal value of the integral, $\xi$ is the integration variable in wavenumber. Eqn. (S3) ensures that the real and imaginary parts of the relative permittivity are Kramers-Kronig consistent. The resulting fitting parameters, including the thickness, are shown in Table S1. This model fits the experimental ellipsometry data (500 – 1200 nm) well (as shown in Fig. S1a-b).

We then obtain the complex refractive index by summing the contributions of all the Gaussian oscillators and $\epsilon_\infty$, which is the value of the relative permittivity at frequencies much higher than the highest-frequency oscillator, and is also a fitting parameter (the fitted value of $\epsilon_\infty$ is also listed in Table S1):

$$(n + i\kappa)^2 = \epsilon_\infty + \sum_{m=1}^{7} \epsilon_{r,m}, \tag{S5}$$

where $n$ is the real part of the refractive index and $\kappa$ is the extinction coefficient (Fig. S1c).

**Table S1.** Fitting parameter values of the oscillator model for the Y6/rubrene/DBP BHJ sample

| Oscillator No. | $A$ | $E_0$ (eV) | $Br$ (eV) | $\epsilon_\infty$ | Thickness (nm) |
|---|---|---|---|---|---|
| 1 | 0.6635 | 1.4956 | 0.1409 | | |
| 2 | 0.6523 | 1.6253 | 0.2527 | | |
| 3 | 0.5294 | 1.8752 | 0.4347 | | |
| 4 | 0.2254 | 2.3067 | 0.1098 | 1.3187 | 98.83 |
| 5 | 0.6178 | 2.4159 | 0.4826 | | |
| 6 | 0.3701 | 3.0750 | 0.6497 | | |
| 7 | 16.077 | 6.6958 | 0.8268 | | |

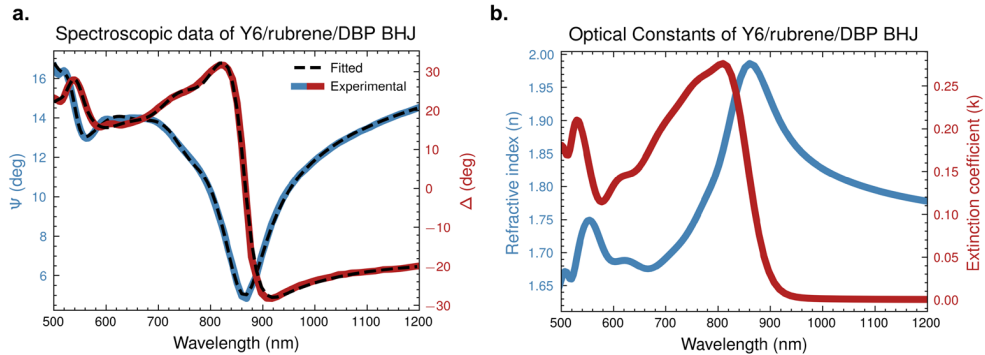

**Fig. S1: Characterization of optical constants of Y6/rubrene/DBP BHJ via ellipsometry**. **a**. Ψ and Δ data from the ellipsometry measurement. **b**. Fitted optical constants of the Y6/rubrene/DBP BHJ.

We repeated the measurements and fitting processes for different Y6/rubrene/DBP BHJ samples with the same preparation recipe as discussed in Methods section, and we found that the oscillator model in Table S1 fits all those samples well.

## S2. Light extraction enhancement from the dichroic backreflector

To model the light extraction enhancement due to the addition of the dichroic backreflector coating we used an electric-dipole source in a 3D finite-difference time-domain (FDTD) simulation to model fluorescence in the upconverting material. This is done by running three separate simulations with a single dipole oriented in the $x$, $y$, and $z$ directions. The far-field electric field data is measured in all three simulations as shown in Fig. S2a, converted to radiated power, and then all three results are averaged out to approximately model an isotropic emitter in the Y6/rubrene/DBP BHJ. To make the model better represent emission from throughout the BHJ the calculation is repeated for several dipole locations along the 100 nm thickness of the upconverting layer, and the results are averaged.

This process is repeated without a dichroic backreflector in the simulation (i.e., using just a glass substrate), and then we divide the average far-field power for these two cases (with the backreflector divided by without the backreflector) to get the light extraction enhancement vs. wavelength plot shown in Fig. 2b in main text. To illustrate the benefit of adding the backreflector, we also show the far-field patterns (wavelength vs angle) for the 2D FDTD simulation in Fig. S2b-c.

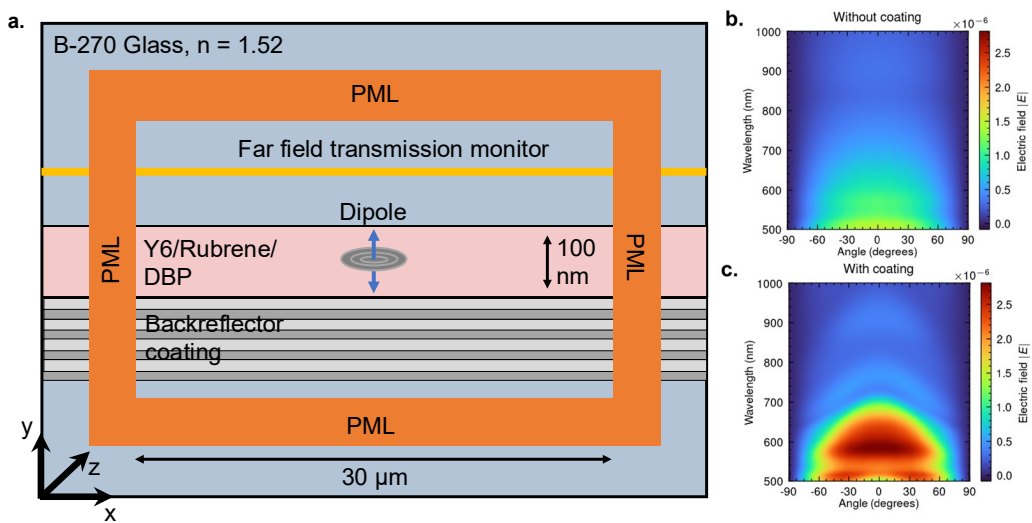

**Fig. S2: Light extraction enhancement simulations. a**, 3D FDTD simulation used to measure light extraction enhancement with the upconverter and our dichroic backreflector thin-film coating. The far-field transmission monitor performs far-field projection and applies Snell's law to account for transmission in air. **b-c,** The far field emission pattern from the 2D FDTD simulation, with and without the dichroic backreflector coating.

## S3. Lateral beam shift from the dichroic backreflector

The dichroic backreflector puts us at risk of degrading imaging quality due to lateral beam shifts[3] at the backreflector/BHJ interface. To calculate potential resolution loss due to reflections from the dichroic backreflector, we set up the 2D FDTD simulation shown in Fig. S3a. A Gaussian beam with 1 μm beam width and wavelength 610 nm, is incident at 20 degrees from Y6/rubrene/DBP onto the dichroic backreflector on a glass substrate. We run this simulation and monitor the beam shift (as compared to a perfect metal mirror) due to the backreflector coating. This beam shift turns out to be 0.32 μm for p-polarized light, and 0.13 μm for s-polarized light. Both of these values are below the current imaging resolution and the ideal diffraction limited resolution for our system. At angles greater than 20°, we start approaching Brewster's angle for the whole structure. Hence the reflection for p-polarized light starts dropping until all p-polarized light is transmitted at ~34°, while the s-polarized light is still fully reflected. In addition, after around 38°, all light emitted from Y6/rubrene/DBP layer is totally internally reflected at the glass-air interface, setting the angle of the escaping light cone from the upconverter.

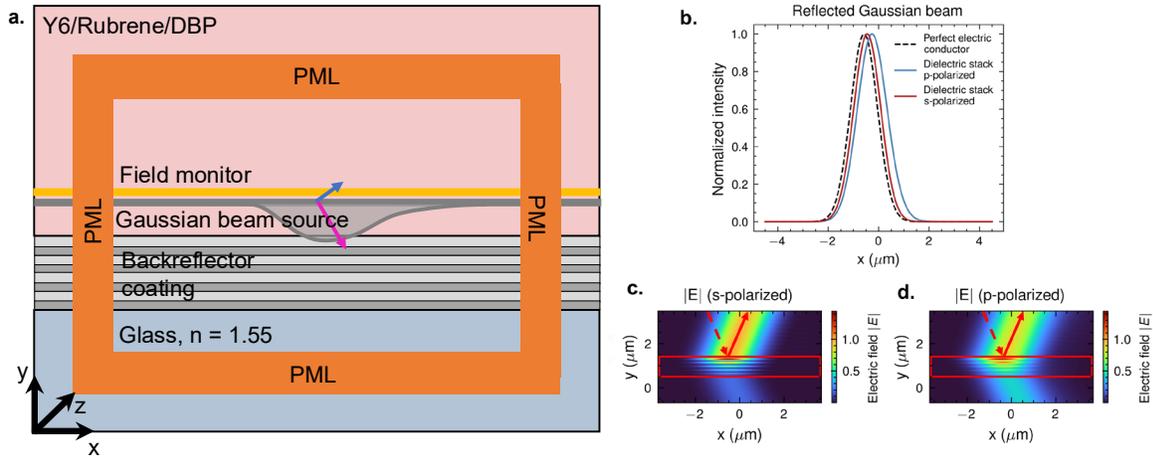

**Fig. S3: Simulations of the beam shift from the dichroic backreflector. a**, 2D FDTD simulation at the interface between the BHJ and the dichroic backreflector coating. A tilted Gaussian beam is incident on the thin film coating and a field monitor is used to measure the reflected beam at the imaging wavelength. **b,** Spatial profile of the reflected beam for s and p-polarized input with reference to the perfectly reflected beam. **c-d,** XY field profile for s and p-polarization. The dichroic backreflector is outlined with a red box. The incident and reflected beam are indicated with a dashed and solid red arrow respectively.

## S4. Absorption enhancement due to the gold nanopillars

To calculate the absorption due to gold pillars embedded in the upconverting material, we set up and run the 3D FDTD simulation shown in Fig. S4a-b. We measure the absorption inside Y6/rubrene/DBP BHJ by first separating the grid inside the absorption monitor based on the material index, and then integrating over all grid points which correspond to the BHJ.

To calculate of expected enhancement using gold pillars with the near-infrared (NIR) LED (Thorlabs, MBB2L1), we multiply the absorption vs. wavelength (Fig. 3b) with the spectrum of the LED, integrate over the 700 – 1000 nm range, and divide by the integral for the bare BHJ, to obtain the absorption enhancement of ~2.1 with this LED source.

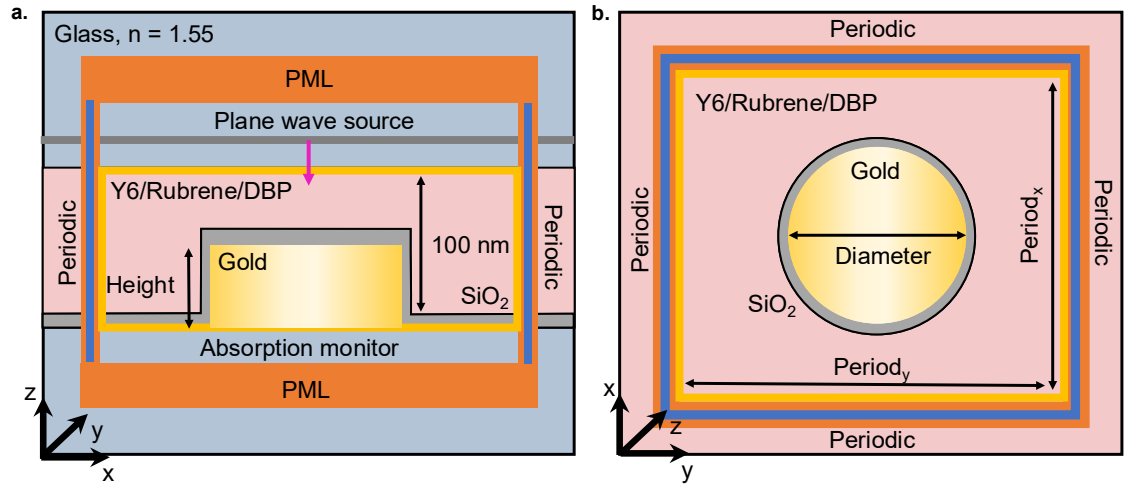

**Fig. S4: FDTD simulation of the absorption enhancement due to plasmonic nanopillars. a-b**, XZ (side) and XY (top) profile of the FDTD simulation used to measure the absorption enhancement inside the BHJ due to embedded gold nanopillars.

## S5. Zemax raytracing of lenses around the upconverter for minimum intensity measurement:

Fig. S5 shows the lens setup used for the minimum intensity measurement results shown in Fig. 4b. This lens setup is consistent with Fig. 2e power measurement setup, the only difference is the higher degree of focusing (controlled by positioning of L1 and L2) and resulting smaller beam spot size on the upconverter. We use an achromatic doublet (L1, Fig. 2e) and aspheric lens (L2, Fig. 2e) to focus a wide NIR beam on the upconverter, down to a 0.38 mm$^2$. After the upconverter, a configuration of three singlet lenses (L3-L5, Fig. 2e) is used to capture light from the upconverter and focus it down to a 0.47 mm$^2$ beam spot on the power meter.

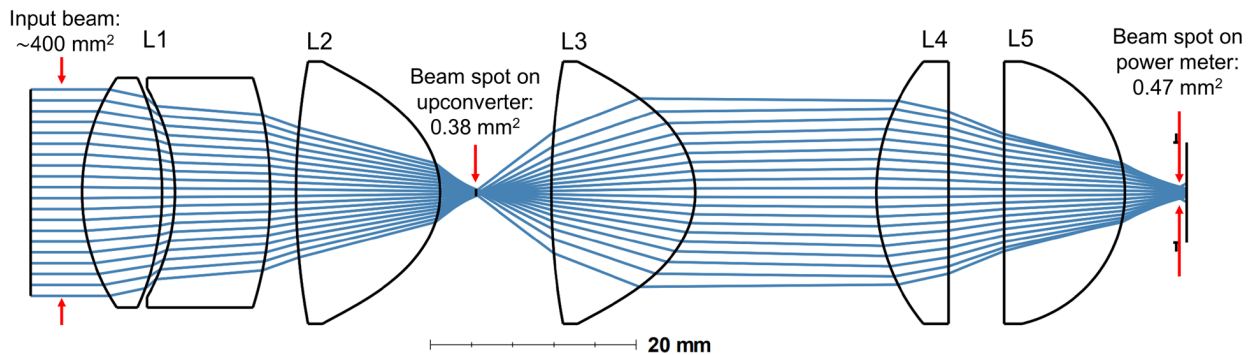

**Fig. S5: Zemax simulations.** Zemax raytracing of the lens configuration used to focus light on the upconverter, and then capture emitted light and focus it on the power meter. All rays correspond to light at a wavelength of 610 nm.

## S6. Proposed imaging system for human eye viewing, and calculations:

Fig. S6 shows the proposed optical setup to be used for human eye viewing, in conjunction with Supp. S5 and Fig. 4b results. Lenses L4 and L5 in Fig. S5 are replaced with a beam reducer to reduce the collimated beam to the width of the human eye pupil, which is approximately 8 mm in diameter (fully dilated, operating under dark-adapted vision). Since we measure approximately ~0.01 nW of upconverted light at the lowest input intensity on the power meter (Supp. S5, Fig. 4b), we expect that a collimated beam of 0.5 cm$^2$ area to have an irradiance of 0.02 nW/cm$^2$ on the pupil.

The absolute threshold for scotopic (dark-adapted) vision is $10^{-6}$ lm/m$^2$ [4,5]. For the peak wavelength in our system, i.e., 610 nm, the luminous efficacy is 27 lm/W. Using these two values, we arrive at a beam intensity requirement of nearly 0.005 nW/cm$^2$ for an incident collimated beam with a diameter which is equal to or less than the pupil diameter. This is nearly an order of magnitude below the irradiance we would expect from our proposed imaging setup for human eye viewing. Hence, we can expect to detect visible intensities by a human eye with our upconversion imaging system operating at an input of intensity of ~50 nW/cm$^2$.

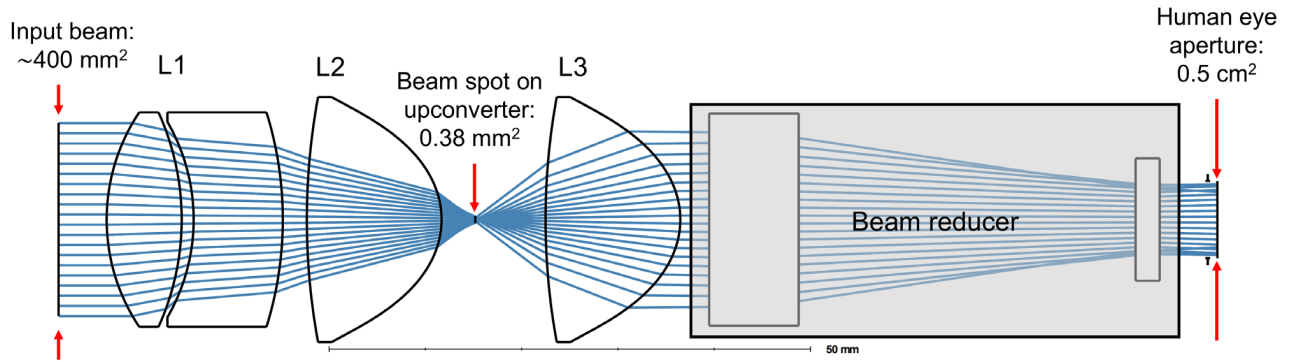

**Fig. S6: Optical system layout for proposed human eye viewing.** Zemax raytracing of the proposed optical system for human eye viewing. All rays correspond to light at a wavelength of 610 nm.

## S7. Reduction of the quadratic-to-linear threshold in the plasmonic sample, and the effect of passivation on upconversion yield

Figures S7a-e show the log-log plots of the input NIR intensity vs. the output visible power within and outside the region with gold nanostructures, for 5 separate upconverting devices. Instead of observing a sharp kink at the threshold point[6,7], we observe a curve. This is likely because the input wavelength range is broad (~150 nm), hence, near the threshold value, there is a mix of photons in quadratic region versus linear region, leading to more smoothed over kink at the threshold point. However, even with this smoothing, it is very easy to observe that the quadratic-to-linear threshold within the region with gold nanopillars compared to the region without them.

We also show the electronic quenching of visible output without the presence of a passivation layer between the Y6/rubrene BHJ and gold nanopillars. We measure only a 1.5 factor of increase in slope with non-passivated gold pillars. We found that atomic layer deposition (ALD) of aluminum oxide (Al$_2$O$_3$) is not effective at preventing quenching. We expect that this is due to incomplete passivation (pinhole defects) as gold is hydrophobic and hence the $Al_2O_3$ does not nucleate well[8,9]. Instead, we used plasma-enhanced chemical vapor deposition (PECVD) to deposit a passivation layer of silicon dioxide ($SiO_2$) atop the gold nanopillars. After depositing 7 nm of $SiO_2$ through PECVD, we measure a 2.2 enhancement, as we expected from simulations that do not assume any electronic quenching.

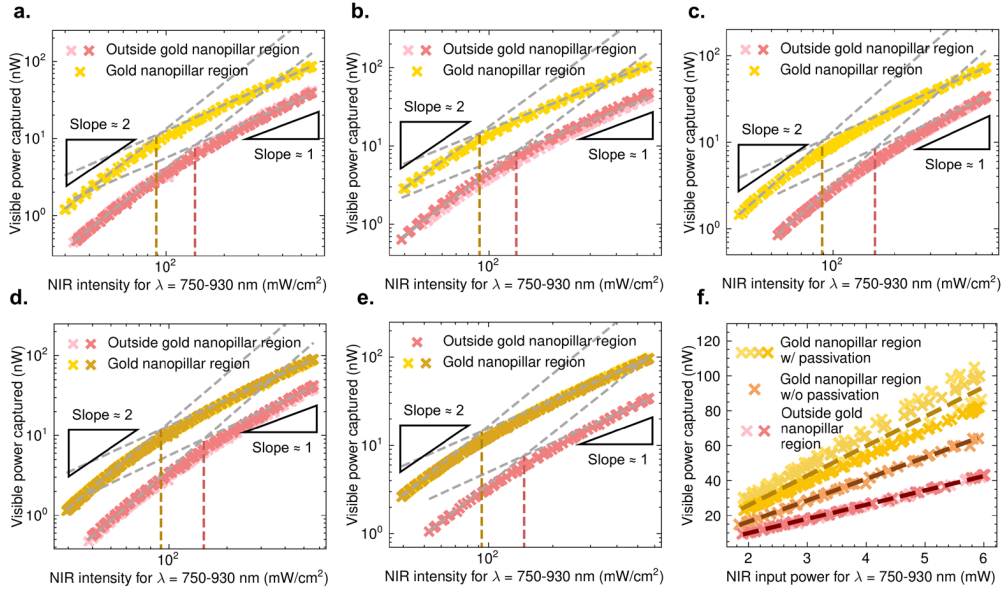

**Fig. S7: Threshold with gold pillars and the effect of passivation. a-e**, Log-log plots showing the quadratic-linear threshold for our upconversion devices with gold nanopillar regions. The vertical dashed pink and yellow lines approximately show the location of the threshold for outside and inside the gold region specifically. These threshold values are given by: a: 88 mW/cm$^2$ and 140 mW/cm$^2$, b: 90 mW/cm$^2$ and 134 mW/cm$^2$, c: 88 mW/cm$^2$ and 160 mW/cm$^2$, d: 90 mW/cm$^2$ and 150 mW/cm$^2$, e: 93 mW/cm$^2$ and 147 mW/cm$^2$. **f,** Measured visible power from a non-passivated gold nanopillar upconverter, compared with the passivated upconverters, demonstrating the quenching of visible output with the absence of a passivation layer

## S8. Video of translation of square array of gold nanopillars in transmission imaging setup

This video (attached separately; screenshot below) demonstrates the increase in brightness on the upconverter within the gold region. We translate an upconverter with a gold patterned square with grid size (2 mm)$^2$ on an XY translation stage, while using the transmission imaging setup shown in Fig. 1f. This video is captured using the same settings described in transmission imaging setup, at 60 frames per second.

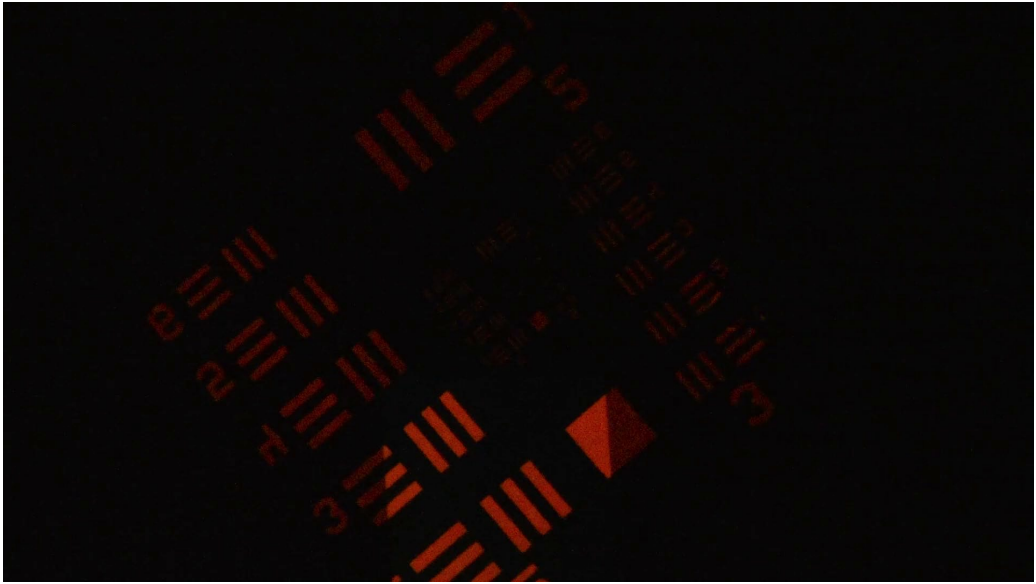

**Fig. S8: Screenshot of video showing translation of square gold nanopillar array.** Video showing the translation of gold nanopillar patterned upconverter in the transmission imaging setup.

## S9. Control and FDTD simulations for the fully integrated upconverter

The preparation of the fully integrated upconverter is slightly different from other upconverting samples. We found that due to airgaps between the two glass pieces, we needed to spin on both sides (backreflector and gold nanopillars) and then press down and encapsulate them together for the enhancements on both sides to work. Specifically, we coat half the thickness (~50 nm) on the backreflector side, and full thickness (~100 nm) on the gold array side. We arrived at the optimal preparation method for the fully integrated upconverter empirically. We found that spinning full concentration on both sides leads to quenching of emitted light and hence lower enhancement. On the other hand, spinning half concentration on both sides is also detrimental to the functioning of the gold nanopillars, since they are not submerged fully in the BHJ.

For fair comparison, we also prepared a control upconverter for the fully integrated upconverter. This upconverter was prepared by spin coating full concentration of the Y6/rubrene/DBP mixture on one glass substrate, and spin coating half concentration on another glass substrate. These coated glass substrates are allowed to anneal separately and then they are clamped together and encapsulated using epoxy. Fig. S9 shows the visible power measured from this control sample, compared to the bare BHJ upconverter, which is prepared normally. We expect that the increase in upconversion yield by a factor of ~1.6 is due to absorption from a total thicker film, plus some cavity effects due to air gap between the two layers of Y6/rubrene/DBP inside the control upconverter.

We also calculate total enhancement of the fully integrated upconverter using separate FDTD simulation setups: 1) Absorption enhancement (in Supp. S4) and 2) Emission enhancement (in Supp. S2). We multiply the enhancement for these separate simulations in order to estimate the total enhancement from the structure. Note that in the emission enhancement FDTD setup, we use a 4 x 4 μm simulation in *x-z* plane (to reduce simulation size). In a single *x-z* plane, we simulate dipoles in different orientations and locations in order to fully sample one unit cell of the simulation, and then repeat several times for different *y* values across the thickness of the Y6/rubrene/BHJ. Assuming the two layers of upconverting material are in direct contact and that the absorption and emission enhancements multiply, we expected we would get nearly ~4.6-fold enhancement. However, the presence of an air gap decreases the enhancement exponentially, as illustrated by the simulated values plotted in Fig. S9c. Since we experimentally measure an enhancement of 3.9, we expect that the air gap size in our fully integrated upconverter is nearly 0.25 μm.

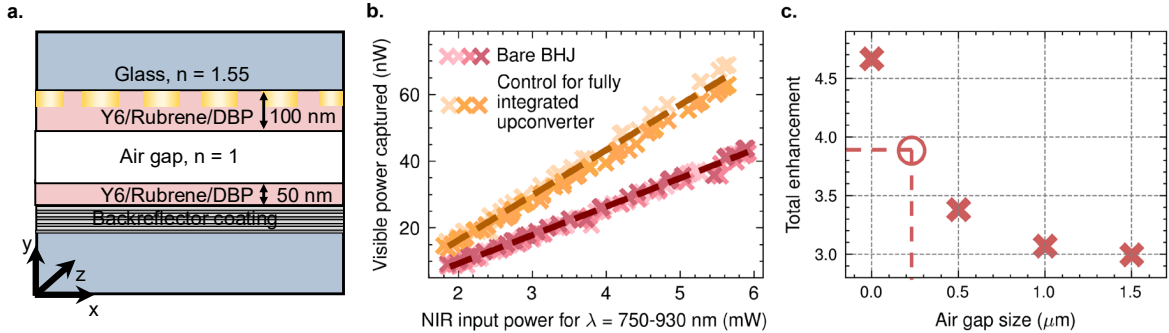

**Fig. S9: FDTD simulations for fully integrated upconverter with varying air gap size. a**, Schematic of the fully integrated upconverter with an air gap between the two layers of Y6/rubrene/DBP BHJ. **b**, Plot showing the measured visible power against the input NIR power to the control upconverter, with the bare BHJ data plotted as reference. **c**, The total enhancement calculated using FDTD simulations for varying air gap size. We expect that the air gap size was in the order of ~0.25 μm (highlighted by circle) from the experimentally measured enhancement and calculated trend.

## S10. Resolution analysis and calculation for fully integrated upconverter

To characterize the limiting resolution of our upconverting system, we analyze the Air Force target images taken using the transmission imaging setup in Fig. 1f. We measure the contrast (or percentage difference) across the line pairs and set our threshold to be 30%. Using this threshold, we observe that we can resolve down to group 6, element 6 (114 lp/mm), but are just shy of resolving group 7, element 1 (128 lp/mm). Finally, to account for magnification of the target on the upconverter, we divide the resolution (in lp/mm) by the magnification (in this case, 1.17) to calculate the true resolution achieved on the upconverter. This gives us a range of ~100 – 110 lp/mm for the resolution achieved on the upconverter. This range corresponds well with the expected 14 μm focal length shift of L1 (NIR achromatic relay lens, from Fig. 1f) across the absorbed NIR light (750 – 930 nm) for this upconverting system, which sets the resolution limit.

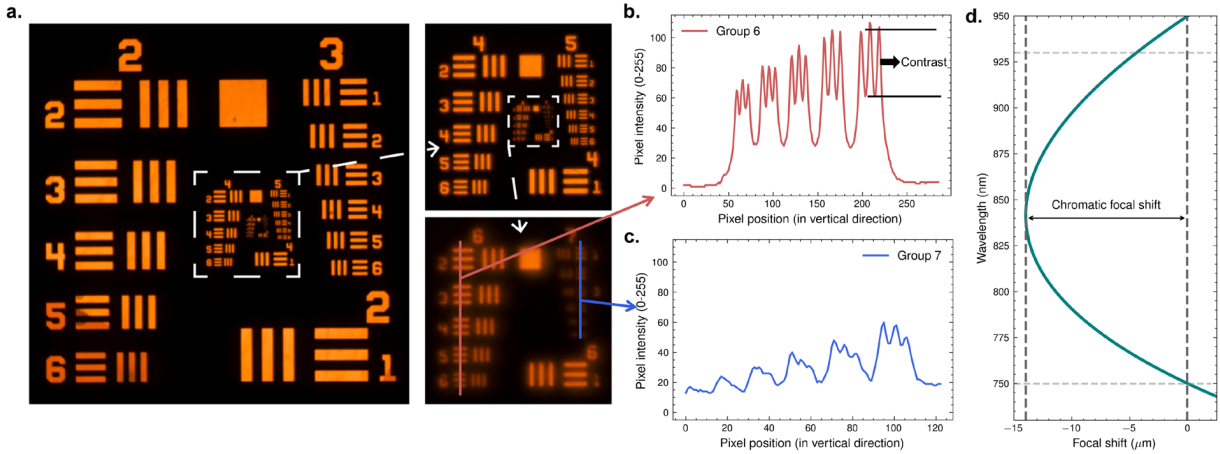

**Fig. S10: Resolution analysis using transmission mode imaging of Air Force target. a**, Air Force target imaged using the setup shown in Fig. 1f with the fully integrated upconverter. **b-c,** Line plots across the horizontal line pairs in group 6 (top) and group 7 (bottom). **d**, Chromatic focal length shift for the NIR achromatic relay lens (L1) used calculated using the Zemax lens file.

## S11. A simplified comparison of input power intensities required for other solid-state frequency upconversion systems in the NIR

It is difficult to compare metrics such as efficiency across different upconverting systems due to the various detection methods, different working wavelength ranges, and external sources used. Instead, we compare at a high level the lowest working input power intensity for several different upconverting systems in the NIR.

Non-linear frequency mixing techniques such as second harmonic generation and sum frequency generation generally require very high intensity pulsed pump source to upconvert efficiently. In the case of second harmonic generation – a single high powered source acts as the pump and the signal beam – allowing upconversion to take place at the scale of 10s of MW/cm$^2$ intensities[10,11], going down to 0.1 MW/cm$^2$ for upconversion imaging from 1064 nm to 532 nm with the use of cavities[12]. Sum frequency generation requires a separate signal and the pump source, and while the signal beam intensity can be lowered by several magnitudes, the pump source still needs to be high intensity to upconvert. Shih et al recently demonstrated a transparent organic upconverting device which can upconvert low NIR (940 nm) power densities (750 nW/cm$^2$) to visible luminescence[13]. However, the caveat to this system is that it requires a voltage source across the upconverting device to function.

Photon upconversion methods such as TTA and lanthanide-doped nanoparticles enable upconversion from weak incoherent light without an external power source. TTA in particular has high efficiency at low power densities (~10 mW/cm$^2$)[6]. Our upconverting system takes this a step forward, using focusing lenses and nanophotonic integrations to reduce the input power density at which we can easily collect (no integrating sphere, only a few lenses, Supp. S5) visible light at human eye sensitivities.

**Table S2.** Minimum input power intensity required for operation of solid-state NIR upconversion systems

| Upconversion system | External power input | Minimum input power density (approx.) |
|---|---|---|
| Second-harmonic generation[11,12] | None | $10^5 - 10^7$ W/cm$^2$ |
| Sum frequency generation[14] | High intensity laser | $0.1 - 1$ W/cm$^2$ |
| Organic upconverting devices[13] | Voltage source | $0.5 - 1$ W/cm$^2$ |
| Lanthanide-doped nanoparticles[15] | None | $0.1 - 1$ W/cm$^2$ |
| Intrinsic triplet-triplet annihilation[6,7] | None | $10^{-3} - 10^{-2}$ W/cm$^2$ |
| Our upconverting system | None | $10^{-7} - 10^{-6}$ W/cm$^2$ |

## S12. Rough estimate of expected system efficiency using previously measured EQE of Y6/rubrene/DBP BHJ:

In this section, as a sanity check, we calculate the system efficiency that we would expect to get, based on previously published external quantum efficiency (EQE) data, and making a number of assumptions about our experimental setup—and then compare against our measured system efficiency.

The EQE of the upconverting heterojunction used in our experiments, without any nanophotonic enhancement (bare BHJ), has been experimentally measured and described in ref. [7]. For the Y6/rubrene/DBP BHJ, the EQE at 100 mW/cm$^2$ is nearly 0.03%. In our transmission-based power measurement setup, only a fraction of emitted light is captured. We use an aspheric lens with 0.8 NA (Thorlabs, ACL25416U-A) which approximately captures a 55° half-angle cone of light (calculated using Zemax). Considering the critical angle within the BHJ (critical angle of ~38°) limit, the half-angle of the escape cone of light in air is ~80°. The addition of our dichroic backreflector, as described in the main text, enhances light collection by a factor of 2.2 by reflecting visible light, and via a small amount of Purcell enhancement (see Section 3.1 in main text). Taking into account this enhancement and solid-angle calculations, we estimate that we are capturing ~47% of emitted light in this setup.

In the main text, we demonstrate that system efficiency can be further enhanced by increasing absorption of NIR light using plasmonic resonators. We experimentally measured an increase of 1.8 in system efficiency in the high intensity (linear) regime (Fig. 4a). In the low intensity (quadratic) regime, we expect this enhancement to increase, as upconversion yield increases (almost) quadratically with input intensity. Using published measurement results from Hu et al[7], we can expect the quadratic region slope to be $\alpha = 1.75$.

At an input intensity of 100 nW/cm$^2$ to the system, we have a corresponding 0.1 mW/cm$^2$ on the upconverter, because the beam spot area is reduced by a factor of 1000. Due to the change in input intensity, we can expect the corresponding EQE of the device to approximately scale with a factor given by: $\left(\frac{0.1 \frac{mW}{cm^2}}{100 \frac{mW}{cm^2}}\right)^\beta$, where $\beta = \alpha - 1$. Combining all these facts, we can expect the system efficiency of the setup in Fig. 4b with the fully integrated device to be given by:

Expected system efficiency at 0.1 mW/cm²
$= $ (EQE of bare BHJ at 100 mW/cm²) × (% Emitted light captured)
× (Purcell enhancement) × (Plasmonic enhancement)$^\alpha$
× (EQE scaling at different intensity)

Which comes out to be:

$$0.03\% \times 0.47 \times 1.2 \times 1.8^{1.75} \times \left(\frac{0.1 \frac{mW}{cm^2}}{100 \frac{mW}{cm^2}}\right)^{1.75-1} = 0.0003\%$$

By comparison, our measured system efficiency at intensity comes out to be ~0.003% (Fig. 4b, main text). This is off from our expected system efficiency by an order of magnitude. We believe that one order of magnitude is reasonable given several differences between the setups used to measure the EQE in Hu et al and the present paper, plus a number of uncertainties in the calculation.

In particular, the EQE results measured by Hu et al[7] used an 808 nm laser, whereas we are using a broadband LED from ~750 nm to ~900 nm. Also, it is likely that the effective quadratic region slope is different due to the structure of the fully integrated upconverter and the broadband light source. Furthermore, it is unlikely that the upconverted intensity remains exactly quadratic with respect to input intensity between 0.1 mW/cm² and at 100 mW/cm². In fact, for intensities close to 100 mW/cm², $\alpha$ is likely greater than 1.75. Additionally, some of the differences could also be attributed to uncertainty errors in beam spot area calculations, and to uncertainty in the input power (see *Methods* on the details of the input power measurement).

## S13. Estimation of nightglow irradiance:

Estimating nightglow irradiance is challenging due to significantly large variability caused by factors such as altitude, cloud cover, humidity, and geographic location[16]. To calculate a rough estimate for nightglow irradiance, we reference a technical report by Vatsia et al[16], which presents spectral radiant sterance measurements of the night sky from 450 to 2000 nm using a ground-based 3-channel Fourier spectroradiometer. From this report, we specifically use data collected in the absence of the moon to measure nightglow only radiance (see Fig. S11). We compute radiance $L$ by numerically integrating over 700–930 nm (working wavelength range for Y6/rubrene/DBP). We then convert radiance $L$ (in nW/cm²/sr) to irradiance $I$ (in nW/cm²) by integrating over the hemisphere above the observation point:

$$I = \int_0^{\phi=2\pi} \int_0^{\theta=\pi/2} L(\theta, \phi) \cdot \cos\theta \cdot \sin\theta \, d\theta d\phi$$

Assuming uniform radiance, as is the case with diffuse nightglow, this simplifies to $I = L\pi$. Hence nightglow irradiance in the NIR regime of 700 – 930 nm is estimated to be ~4 nW/cm$^2$. Using the ESO SkyCalc[17,18] model for atmospheric emission, we simulated this irradiance to be ~2 nW/cm$^2$, which is very close to the value estimated from Vatsia et al.

As a reminder, in the main text we observe upconversion down to at least 50 nW/cm$^2$ for wavelengths 750 – 930 nm (Fig. 4b). Both of these irradiance numbers (4 nW/cm$^2$ for nightglow within the 700-930 nm range and 50 nW/cm$^2$ in our measurement) have a lot of uncertainty, but they are currently our best estimates. We therefore state in the main text that we measure upconversion at irradiances roughly within an order of magnitude of the nightglow irradiance.

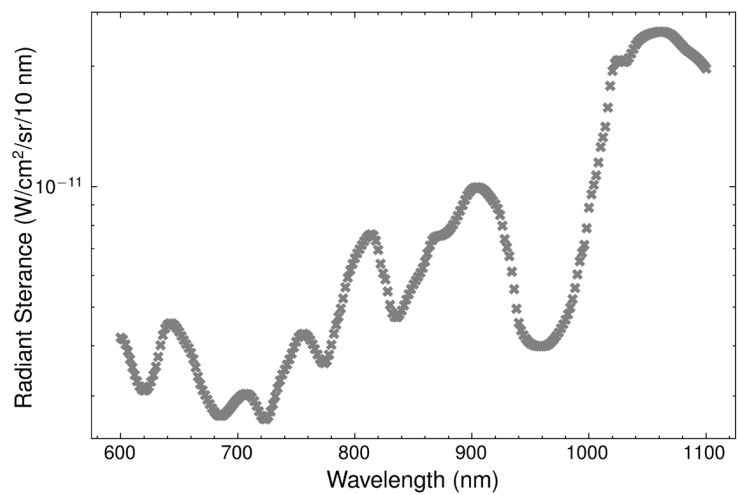

**Fig. S11: Night-sky spectral radiant sterance in the absence of moon.** Figure adapted from ref [16], page number 31.

## Supporting references: